\documentclass[aps,prl,reprint,showpacs,superscriptaddress]{revtex4-1}

\usepackage{graphicx}
\usepackage{dcolumn}
\usepackage{bm}
\usepackage{color}
\usepackage{lipsum}
\usepackage{lineno}
\usepackage[colorlinks,urlcolor=blue,linkcolor=blue,citecolor=blue]{hyperref}

\usepackage{siunitx}
\begin{document}

\title{Anomalous stopping of laser-accelerated intense proton beam in dense ionized matter}%

\author{Jieru Ren}
\affiliation{MOE Key Laboratory for Nonequilibrium Synthesis and Modulation of Condensed Matter, School of Science,
Xi'an Jiaotong University, Xi'an 710049, China}
\author{Zhigang Deng}
\affiliation{Science and Technology on Plasma Physics Laboratory, Laser Fusion Research Center, China Academy of Engineering Physics, Mianyang 621900, China}
\author{Wei Qi}
\affiliation{Science and Technology on Plasma Physics Laboratory, Laser Fusion Research Center, China Academy of Engineering Physics, Mianyang 621900, China}
\author{Benzheng Chen}
\affiliation{MOE Key Laboratory for Nonequilibrium Synthesis and Modulation of Condensed Matter, School of Science,
Xi'an Jiaotong University, Xi'an 710049, China}
\affiliation{Institute for Fusion Theory and Simulation, Department of Physics, Zhejiang University, Hangzhou 310058, China}
\author{Bubo Ma}
\affiliation{MOE Key Laboratory for Nonequilibrium Synthesis and Modulation of Condensed Matter, School of Science,
Xi'an Jiaotong University, Xi'an 710049, China}
\author{Xing Wang}
\affiliation{MOE Key Laboratory for Nonequilibrium Synthesis and Modulation of Condensed Matter, School of Science,
Xi'an Jiaotong University, Xi'an 710049, China}
\author{Shuai Yin}
\affiliation{MOE Key Laboratory for Nonequilibrium Synthesis and Modulation of Condensed Matter, School of Science,
Xi'an Jiaotong University, Xi'an 710049, China}
\author{Jianhua Feng}
\affiliation{MOE Key Laboratory for Nonequilibrium Synthesis and Modulation of Condensed Matter, School of Science,
Xi'an Jiaotong University, Xi'an 710049, China}
\author{Wei Liu}
\affiliation{MOE Key Laboratory for Nonequilibrium Synthesis and Modulation of Condensed Matter, School of Science,
Xi'an Jiaotong University, Xi'an 710049, China}
\affiliation{Xi'an Technological University, Xi'an 710021, China} 
\author{Dieter H.H. Hoffmann}
\affiliation{MOE Key Laboratory for Nonequilibrium Synthesis and Modulation of Condensed Matter, School of Science,
Xi'an Jiaotong University, Xi'an 710049, China}
\author{Shaoyi Wang}
\affiliation{Science and Technology on Plasma Physics Laboratory, Laser Fusion Research Center, China Academy of Engineering Physics, Mianyang 621900, China}
\author{Quanping Fan}
\affiliation{Science and Technology on Plasma Physics Laboratory, Laser Fusion Research Center, China Academy of Engineering Physics, Mianyang 621900, China}
\author{Bo Cui}
\affiliation{Science and Technology on Plasma Physics Laboratory, Laser Fusion Research Center, China Academy of Engineering Physics, Mianyang 621900, China}
\author{Shukai He}
\affiliation{Science and Technology on Plasma Physics Laboratory, Laser Fusion Research Center, China Academy of Engineering Physics, Mianyang 621900, China}
\author{Zhurong Cao}
\affiliation{Science and Technology on Plasma Physics Laboratory, Laser Fusion Research Center, China Academy of Engineering Physics, Mianyang 621900, China}
\author{Zongqing Zhao}
\affiliation{Science and Technology on Plasma Physics Laboratory, Laser Fusion Research Center, China Academy of Engineering Physics, Mianyang 621900, China}
\author{Leifeng Cao}
\affiliation{Science and Technology on Plasma Physics Laboratory, Laser Fusion Research Center, China Academy of Engineering Physics, Mianyang 621900, China}
\author{Yuqiu Gu}
\affiliation{Science and Technology on Plasma Physics Laboratory, Laser Fusion Research Center, China Academy of Engineering Physics, Mianyang 621900, China}
\author{Shaoping Zhu}
\affiliation{Institute of Applied Physics and Computational Mathematics, Beijing 100094, China}
\affiliation{Science and Technology on Plasma Physics Laboratory, Laser Fusion Research Center, China Academy of Engineering Physics, Mianyang 621900, China}
\affiliation{Graduate School, China Academy of Engineering Physics, Beijing 100088, China} 
\author{Rui Cheng}
\affiliation{Institute of Modern Physics, Chinese Academy of Sciences, Lanzhou 710049, China}
\author{Xianming Zhou}
\affiliation{MOE Key Laboratory for Nonequilibrium Synthesis and Modulation of Condensed Matter, School of Science,
Xi'an Jiaotong University, Xi'an 710049, China}
\affiliation{Xianyang Normal University, Xianyang 712000, China}
\author{Guoqing Xiao}
\affiliation{Institute of Modern Physics, Chinese Academy of Sciences, Lanzhou 710049, China}
\author{Hongwei Zhao}
\affiliation{Institute of Modern Physics, Chinese Academy of Sciences, Lanzhou 710049, China}
\author{Yihang Zhang}
\affiliation{Institute of Physics, Chinese Academy of Sciences, Beijing 100190, China}
\author{Zhe Zhang}
\affiliation{Institute of Physics, Chinese Academy of Sciences, Beijing 100190, China}
\author{Yutong Li}
\affiliation{Institute of Physics, Chinese Academy of Sciences, Beijing 100190, China}
\author{Dong Wu} \email{dwu.phys@zju.edu.cn}
\affiliation{Institute for Fusion Theory and Simulation, Department of Physics, Zhejiang University, Hangzhou 310058, China}
\author{Weimin Zhou} \email{zhouwm@caep.cn}
\affiliation{Science and Technology on Plasma Physics Laboratory, Laser Fusion Research Center, China Academy of Engineering Physics, Mianyang 621900, China}
\author{Yongtao Zhao} \email{zhaoyongtao@xjtu.edu.cn}
\affiliation{MOE Key Laboratory for Nonequilibrium Synthesis and Modulation of Condensed Matter, School of Science,
Xi'an Jiaotong University, Xi'an 710049, China}
\bibliographystyle{apsrev4-1}

\date{\today}

\begin{abstract}

Ultrahigh-intensity lasers (10$^{18}$-10$^{22}$W/cm$^{2}$) have opened up new perspectives in many fields of research and application \cite{rygg2008,gu2018,vranic2018,li2020,kodama2001fi}. By irradiating a thin foil, an ultrahigh accelerating field (10$^{12}$ V/m) can be formed and multi-MeV ions with unprecedentedly high intensity (10$^{10}$A/cm$^2$) in short time scale ($\sim$ps) are produced \cite{la1,la2,cowan2004,roth2002,hegelich2002,hatchett2000,snavely2000,maksimchuk2000,sheng2000}. Such beams provide new options in radiography \cite{cobble2002}, high-yield neutron sources \cite{roth2013}, high-energy-density-matter generation \cite{wdm}, and ion fast ignition \cite{roth2001fast,wang2015}. An accurate understanding of the nonlinear behavior of beam transport in matter is crucial for all these applications. We report here the first experimental evidence of anomalous stopping of a laser-generated high-current proton beam in well-characterized dense ionized matter. The observed stopping power is one order of magnitude higher than single-particle slowing-down theory predictions. We attribute this phenomenon to collective effects where the intense beam drives an decelerating electric field approaching 1GV/m in the dense ionized matter. This finding will have considerable impact on the future path to inertial fusion energy.

\end{abstract}

\maketitle
Alpha particle stopping in dense ionized matter is essential to achieve ignition in inertial confinement fusion \cite{hurricane2016alpha,atzeni2004fusion,paradela2019maglif,betti2016icf,jacquemot2017icf}. Fast ignition (FI) relies even more on a detailed understanding of untrahigh-current ion stopping in matter, which is therefore considered as a fundamental process of utmost importance to nuclear fusion. In the fast ignition scheme \cite{F1,F2}, a short and intense pulse of energetic charged particles - electrons, protons or heavy ions - generated by an ultrahigh intensity laser, is directed towards the pre-compressed fusion pellet. The charged-particle beam requirements to achieve ignition have been discussed and studied in detail previously \cite{fernandez2009progress,F3,F4,F5,F6} based on single-particle stopping theory. 
However, the collective effects induced by high-current charged particle beams could alter significantly the projected range, the magnitude of energy deposition, and therefore change the requirements for ignition correspondingly. Besides, in the cases of ion beam driven inertial confinement fusion and high energy density sciense, which requires the highest beam intensity from accelerators \cite{hofmann2018review,wdmprl,wdm1,wdm2,wdm3},  no collective effects on ion stopping processes due to high beam intensity are considered nor - to the best of our knowledge - were they reported in any previous experiments. 

Since the discovery of alpha decay and the availability of energetic fission fragments, it became interesting to study fast particle stopping processes in matter \cite{ziegler1999stopping}. In past decades, numerous theoretical models \cite{ionstopping1,ionstopping3,ionstopping5,ionstopping6,ionstopping7,ionstopping8,deutsch2018}, some of which can be considered to be further developments of the early work of Bethe \cite{ionstopping5} and Bloch \cite{ionstopping6}, are built to describe single charged particle stopping in dense ionzed matter. Only recently experiments with sufficient precision were carried out with dense ionzied matter to distinguish between different models \cite{frenje2019,CKLi,Cayzac}. In these experiments, incident particles are generated from laser induced nuclear reactions \cite{frenje2019,CKLi}, or from traditional accelerators \cite{Cayzac}. Hence the beam intensity was too low to test the model of single particles interacting with dense ionized matter. 

\begin{figure*}
\includegraphics[width=0.95\linewidth]{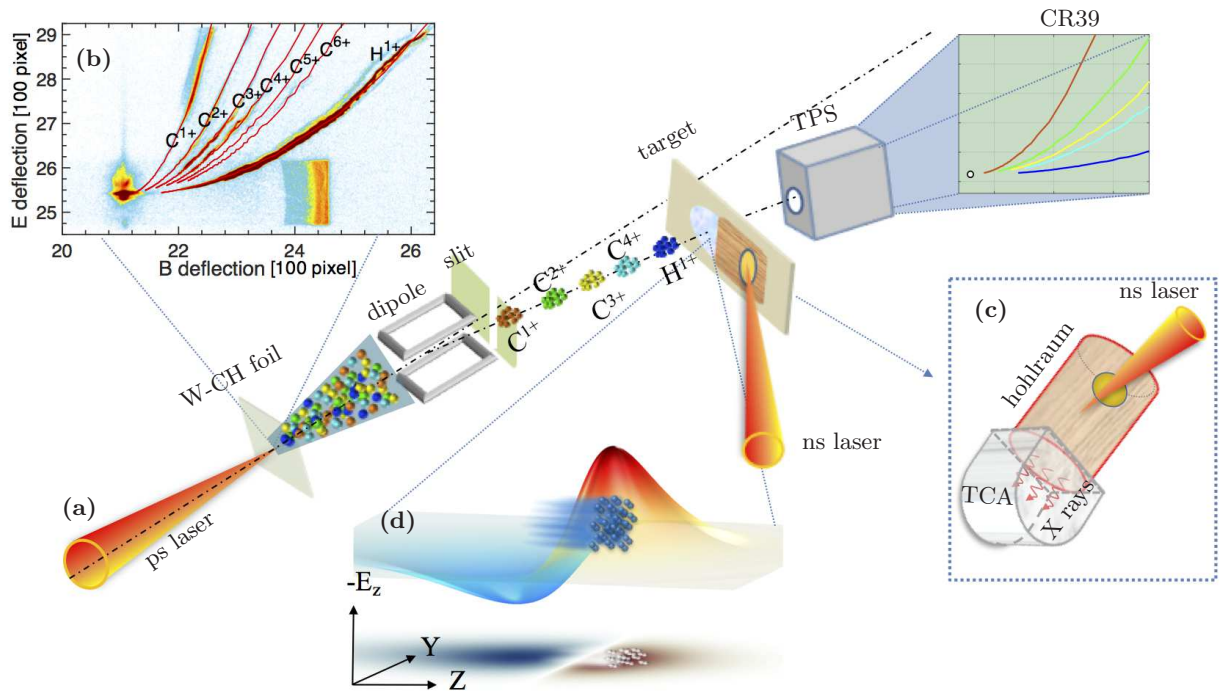}
\put(-460,75){\textbf{(a)}}
\put(-430,55){\rotatebox{35}{ps laser}}
\put(-416,115){\rotatebox{28}{W-CH foil}}
\put(-180,100){ns laser}
\put(-185,220){\rotatebox{15}{TPS}}
\put(-90,260){CR39}
\put(-346,148){\rotatebox{28}{dipole}}
\put(-317,168){\rotatebox{28}{slit}}
\put(-243,209){\rotatebox{28}{target} }               
\put(-455,245){\textbf{(b)}}   
\put(-115,150){\textbf{(c)}}
\put(-340,70){\textbf{(d)}}
\put(-120,80){\rotatebox{357}{TCA}} 
\put(-90,65){\rotatebox{53}{X rays}} 
\put(-105,115){\rotatebox{53}{hohlraum}}
\put(-60,157){{ns laser}}
\caption{\label{fig1} \textbf{Layout of the experiment.} (a) A ps laser is focused onto a tungsten foil, generating intense short-pulse ion beams with different species. A magnetic dipole with slits at entrance and exit serve as $\textbf{p}/q$ analyser to select mono-energetic ion beams. Such ions interact with the laser generated plasma-target and emerge from the target with a lower energy due to the incurred energy loss. The final-state energy is measured by a Thomson parabola in conjunction with CR$39$ film. (b) Parabola spectra of laser-accelerated ions without dipole measured by Thomson parabola in conjunction with Fuji image plate. (c) The target consists of a gold hohlraum converter to produce the soft x-rays that irradiate the TCA foam to generate a dense ionized sample. (d) The insert shows the simulation result of an intense proton beam moving along the z direction, inducing a strong longitudinal electric field, which is counter-directional to the proton beam propagation, causing anomalous stopping.}
\end{figure*}

Intense particle beams generated from interaction of ultrahigh intensity laser with foil open a new realm, where beam-driven complex collective phenomena are expected to occur \cite{intense1,intense2,intense4,intense5,intense6,intense7}. In particular, the stopping power for these intense beam could be orders of magnitude higher than that for individual particles if the beam intensity is high enough \cite{mccorkle1977,kawata2019,deutsch1990,rule1984}. We ever carried out an experiment in previous, and observed a significant enhancement of energy loss for the laser-accelerated intense proton beam (see \cite{zhao} and supplementary material for details). However, it was difficult to conduct quantitive analysis due to the large energy spread of the beam. In order to improve our understanding of these effects, experimental data with high precision are required.

Here in this work, we improves the precision of the measurements through using a magnetic dipole to trim out a quasi-mono-energetic proton beam with energy spread of $\sim$ $6\%$. The dense ionized matter was produced from irradiating a Tri-Cellulose Acetate (TCA) foam sample with soft x-rays from a laser-heated hohlraum. The hydrodynamic timescale of the target is long compared to the proton beam pulse, hence the target can be considered to be quasi static with well-characterized parameters. This is able to quantitatively study the anomalous stopping of intense beam in dense ionized matter. We observed enhanced stopping by one order of magnitude compared to the classical single particle slowing-down theory. We attribute this phenomenon to a strong decelerating electric field induced by the intense proton beam. Our numerical simulation indicate that this collective effect is the primary cause for the anomalous stopping, and it is likely to have a major impact on nuclear fusion scenarios like fast ignition, alpha-particle self heating, as well as ion driven inertial confinement fusion.

\section{Experimental setup and data taking}

The experiment was carried out at the XG-III laser facility of Laser Fusion Research Center in Mianyang. The experimental layout is displayed in Fig.\ \ref{fig1}. Here a short and intense laser beam of $800$ fs duration, 20 $\mu$m focal spot and $150$ J total energy irradiates a CH-coated tungsten foil (15 $\mu$m-thick), generating charged particle beam. The beam consists of a mixture of protons  (H$^{1+}$) and carbon ions with different charge states ( C$^{1+}$, C$^{2+}$, C$^{3+}$ and C$^{4+}$).  They originate at the backside of the target  by means of the target normal sheath acceleration (TNSA). The predominant particle species is H$^{1+}$, because the charge to mass ratio is maximum for this species and is therefore more effectively accelerated than the lower charge-to-mass ratio species of carbon ions. In view that the TNSA mechanism results in a broad range of particle energies, which is not favorable for quantitative analysis of the particle energy loss. A magnetic dipole, with entrance and exit slits, was used to trim the beam to mono-energetic one. The ions, spatially collimated by the 500 $\mu$m entrance slit, are dispersed laterally by the magnetic dipole according to their specific $\textbf{p}/q$ value, where p and q are the particle momentum and charge respectively. A second 500 $\mu$m exit slit,  selects  the quasi-mono-energetic ion pulses. The selected ions consist of different particle species, with similar $\textbf{p}/q$ value, they have, however, different velocities and therefore arrive at the target pulse by pulse with different time delay.

\begin{figure}[h]
\includegraphics[width=0.9\linewidth]{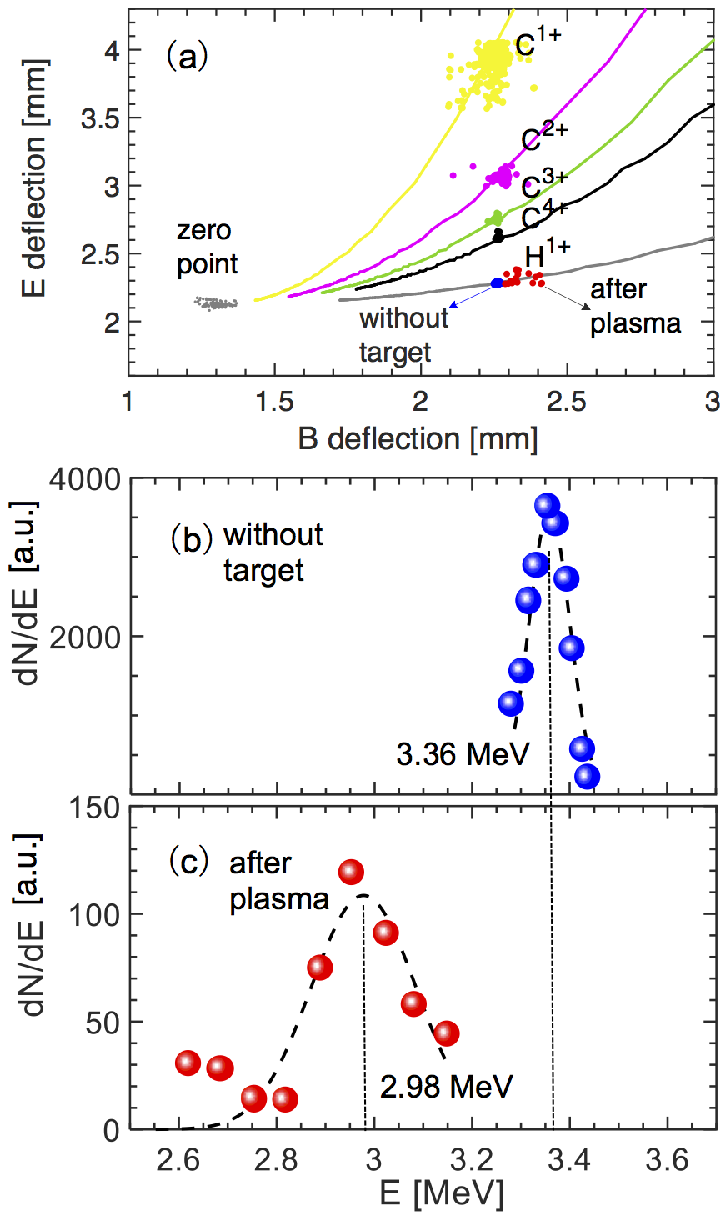}
\caption{\label{fig2} \textbf{TPS CR39 tracks of quasi-monoenergetic ions passing through the system with/without target and the converted energies of protons.} (a) Tracks of ions recorded on TPS CR39 passing through null target and plasma target. The X and Y coordinates represent the magnetic and electric deflection distances.  The dots with the same B deflection from up to down (yellow, magenta, green, black and blue in order) represent tracks formed by C$^{1+}$, C$^{2+}$, C$^{3+}$, C$^{4+}$ and protons passing through the system without target. Tracks for protons passing through plasma are represented by red dots. The tracks for the zero reference point are indicated with grey dots. The dashed curves represent the theoretical tracks of these ion species with various energies. (b) Energy spectra of the protons passing the system without target. The experimental distribution (blue dots) are well fitted by a gaussian profile (dashed line) with central energy of 3.36 MeV (dotted line). (c) Energy spectra of the protons passing the system without target. The experimental distribution (blue dots) are fitted by a gaussian profile (dashed line) with central energy of 2.98 MeV.}
\end{figure}

A gold hohlraum converter was used to generate soft x-ray radiation by interaction of a ns laser pulse [150 J], with the hohlraum walls. The X rays subsequently irradiated and heated the foam target (C$_{9}$H$_{16}$O$_{8}$, density of 2 mg/cm$^{-3}$ and thickness of $1$ mm), into ionized state. Due to the large penetration of soft x-rays, the foam was heated quasi isochorically. The hydrodynamic response of this target material is very well investigated previously \cite{ns1,ns2,ns3}. The hydrodynamic response time is in the ns regime, which is long compared to the proton pulse, hence the target can be considered to be quasi static with well-characterized plasma properties. In order to determine plasma parameters, the emission spectra of the gold hohlraum and target matter were measured. The gold hohlraum radiation spectrum is well represented by a 20 eV black body radiation spectrum, while the temperature of the plasma target is 17 eV. This value was obtained from a Boltzmann slope analysis of the He like carbon lines. Given a temperature of 17 eV,  and mass density of 2 mg/cm$^{-3}$, the number density of free electrons is determined to be $4\times10^{20}$ cm$^{-3}$ based on the FLYCHK code \cite{fly1}.

A thomson parabola spectrometer (TPS) in conjunction with a plastic track detector CR39 was used to obtain the energy spectrum of the charged particles. In Fig.\ \ref{fig2}(a), tracks recorded on CR39  film are displayed for ions passing through the system with/without target.    

When the plasma target is inserted, only protons are observed in the TPS. The deflection distances of protons without/with target are converted to energies in Fig.\ \ref{fig2}(b) and (c), respectively. The incident, unperturbed protons without target appear at 3.36 MeV with a half width of the distribution (FWHM) of 0.06 MeV. Protons passing through the plasma target are downshifted in energy to 2.98 MeV and the FWHM increased to 0.20 MeV. 
\section{Discussion}
\begin{figure}[h]
\includegraphics[width=1\linewidth]{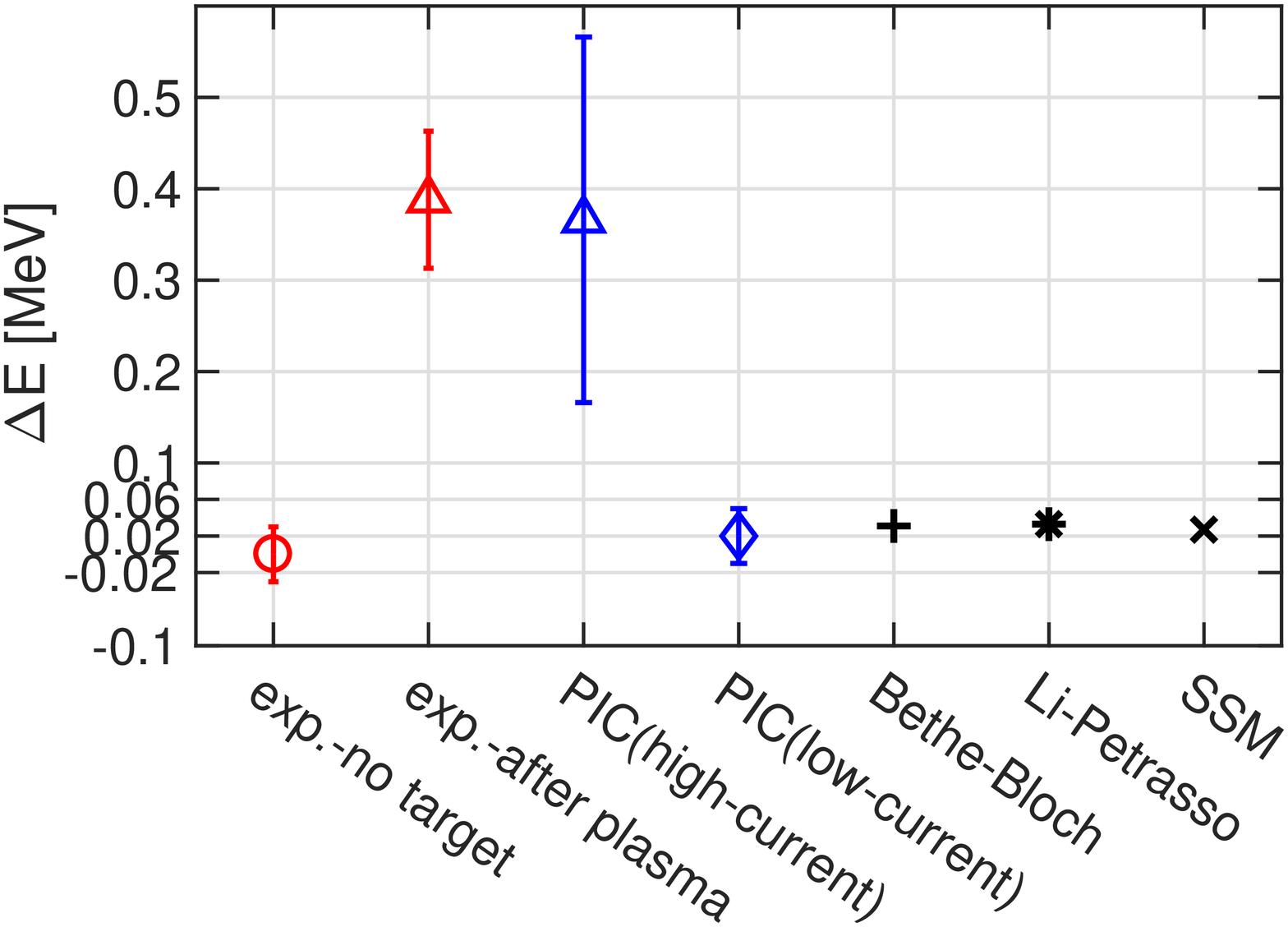}
\caption{\label{fig3} \textbf{Experimental results, numerical (PIC) and analytical prediction of proton energy loss in the dense plasma target (see text for details on plasma parameters)}. The experimental data (red triangle) and PIC simulation data (blue triangle for 3.2 $\times$ 10$^7$ A/cm$^2$ high-current case, blue diamond for 3.2 $\times$ 10$^2$ A/cm$^2$ low-current case) are determined as the central energy downshift, with error bars representing the FWHM of their respective energy spectra. In the analytical calculations, the incident beam energy is assumed to be monoenergetic with the same energy as the central energy of proton beam used in experiment. For comparison, the exprimental data for proton beam energy without target is shown with red circle, but downshifted by its central energy.  }
\end{figure}

In Fig.\ \ref{fig3}, the measured energy loss is compared to theoretical models, e.g. Bethe-Bloch model, Li-Petrasso (LP) theory \cite{ionstopping1} and Standard Stopping Model (SSM) by Deutsch \cite{deutsch2018}. These theories are based on binary collisions with free electrons, bound electrons and/or plasmons. They all underestimate the measured stopping power by as much as one order of magnitude. We therefore call the observed effect anomalous stopping and attribute this to collective electromagnetic effects induced by high-current ion beams.

In order to understand this anomalous stopping, both collective electromagnetic effects and close particle-particle interactions need to be taken into account. The most appropriate tool to simulate the conditions of the experiment is the Particle-in-Cell method (PIC), which in recent years has established itself as a state-of-the-art method for solving problems of kinetic plasma physics. 

\begin{figure}[h]
\includegraphics[width=1\linewidth]{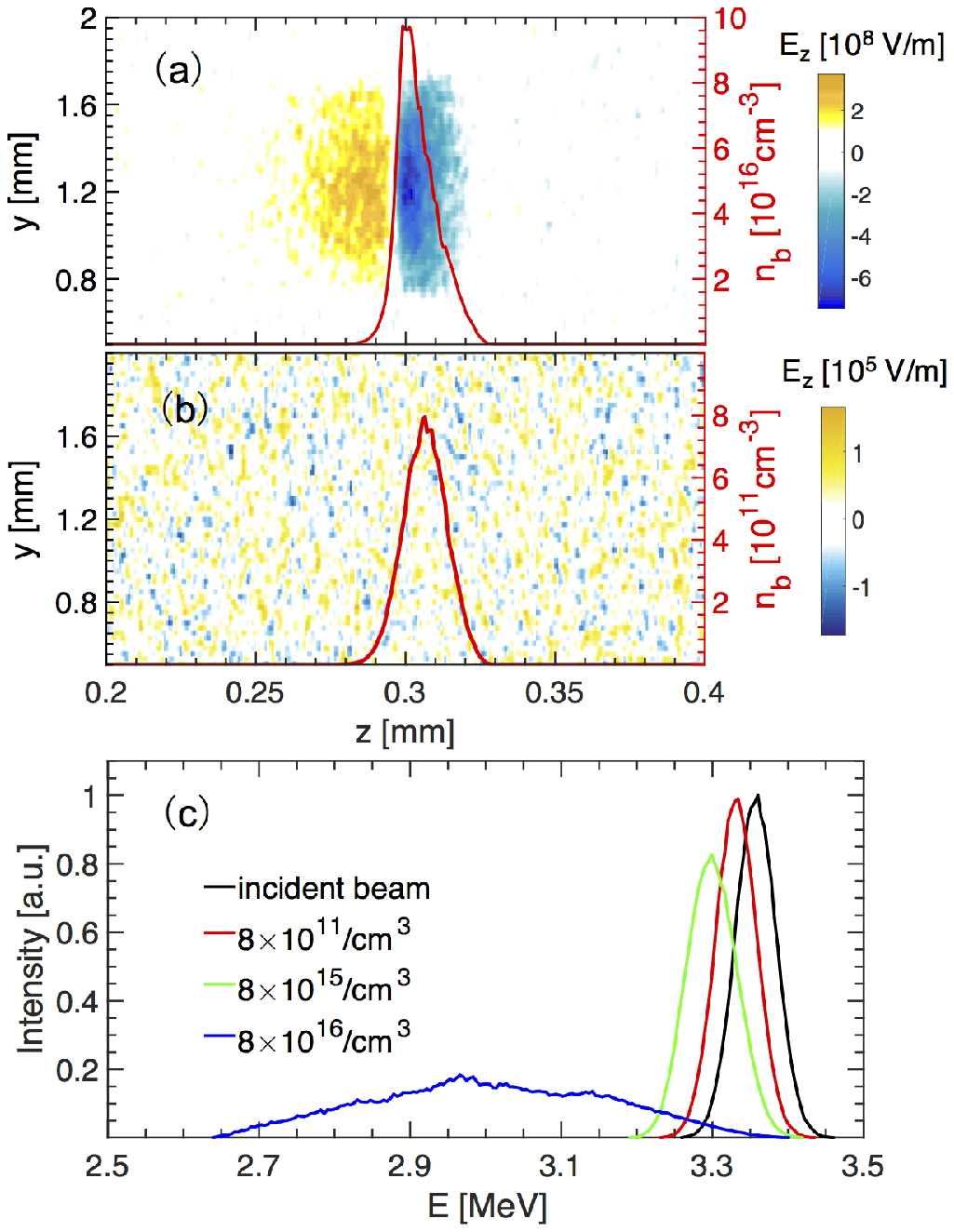}
\caption{\label{fig4} \textbf{Simulated distribution of longitudinal electric field and proton beam density during beam transport in experimentally used ionized target, and the resulted proton beam energy spectra shift after passing through the sample.} (a) Longitudinal electric field driven by proton beam, moving along the Z direction with a beam density of $8\times10^{16}$ cm$^{-3}$. The beam density profile is indicated by the red solid curve. (b) The same situation as in (a) but the initial beam density is reduced to $8\times10^{11}$ cm$^{-3}$. (c) Normalized energy spectra. Incident protons without target interaction are represented by the solid black curve. The colored lines indicate the interaction with the plasma target varying the beam density from $8\times10^{11}$ cm$^{-3}$ (red) to $8\times10^{15}$ cm$^{-3}$ (green), and finally $8\times10^{16}$ cm$^{-3}$ (blue).}
\end{figure}

The simulation assumes, the incident proton beam to have  Gaussian distribution in space and time, with a beam duration of 1 ps and a transverse extension of 1 mm. The energy spectrum is also assumed to be Gaussian, with the peak of the energy distribution at 3.36 MeV and FWHM of 0.06 MeV. The measured ionized target parameters were used as simulation input. The simulation was carried out in Z-Y Cartesian geometry with beam propagating along the Z-direction. The size of the simulation box was  $1.2$ mm $\times$ $2.5$ mm, with a grid size of  $0.75$ $\mu$m $\times$ $25$ $\mu$m, containing 25 particles per grid for each plasma species and 64 particles per grid for the proton beam.

Given the incident proton beam with density of $8\times10^{16}$ cm$^{-3}$, which corresponds to high-current case of $3.2\times10^7$ A/cm$^2$, Fig. \ref{fig4}(a) shows the longitudinal electric field E$_{z}$ induced by the beam-driven return current after a propagation distance of about 0.3 mm. A strong decelerating field approaching $10^{9}$ V/m is generated, and contributes to the proton stopping. The proton energy spectrum after passing through 1 mm of plasma is shown in Fig. \ref{fig4}(c). The energy spread is significantly broadened compared to the initial spread.  We attribute this to a decreasing field, the protons are imbedded in. Protons with higher energies are located at the front end of bunch and therefore experience a smaller decelerating electric field than those with lower energies that come later. The spatial size of this decelerating field is comparable to size of the proton bunch. This is different from the plasma wakefield case \cite{awake1,awake2}, where the spatial structure of the electric field is determined by the plasma density. Here the plasma wakefield wavelength is much smaller than the beam length, therefore the wakefield-induced collective acceleration and deceleration cancel out.
The peak energy of protons is downshifted by 0.39 MeV after passing through the plasma. As shown in Fig. \ref{fig3}, this energy shift (blue triangle) agrees with experimental data in magnitude. 
We carried out additional simulations for different beam densities at $8\times10^{11}$ cm$^{-3}$ and $8\times10^{15}$ cm$^{-3}$, which are defined as low- and intermediate-current cases, respectively. For the low-current case, the beam induced longitudinal electric field $E_z$ after propagating for $0.3$ mm in the plasma is shown in Fig.\ \ref{fig4}(b). No collective decelerating field is excited under such conditions. After passing through the plasma, the energy spectrum is downshifted by only 0.02 MeV as shown in Fig. \ref{fig4}(c).  This prediction agrees well with those calculated by the different binary collision theories, as shown in Fig. 3., which indicates the dominant role of collisional stopping in low-current cases.
As for the intermediate case, the stopping due to the collective effects are comparable to that caused by binary collisions, giving rise to an energy loss of 0.04 MeV as shown in Fig.\ \ref{fig4}(c).

Therefore, the energy loss of laser-accelerated proton beam in the current dense ionized matter is composed of two terms as dE/dx=(dE/dx)$_{collision}$+(dE/dx)$_{collective}$. The first term (dE/dx)$_{collision}$ describes the collisional stopping induced by atomic binary interaction of the individual projectiles with the individual particles in the plasma, which is well predicted by the classic Bethe-Bloch equation and other traditional binary collision theories. The second term (dE/dx)$_{collective}$ describes the collective stopping induced by the self-formed deceleration field, that occurs when sending a very dense ion bunch into the plasma.  Since the ion bunch is imbedded in the deceleration field that increases with the increasing ion bunch density, we expect a significant enhancement of the energy loss (here measured as 20 fold) in the plasma for a very dense ion bunch compared with individual particles incidence. 
\section{Conclusion}
In summary, the laser-accelerated intense proton beam stopping in a dense ionized matter has been measured. Benefiting from the fact that we have a quasi-monoenergetic proton beam and long-living well-characterized dense ionized target, accurate stopping power data were obtained. The measured stopping power exceeds the classical theory predicts in binary collision scheme by about one order of magnitude. The anomalous phenomena can be very well explained by our PIC simulation combined with a new Monte Carlo binary collision model and a reduced model taking account the collective electromagnetic effects. The stopping power is dramatically enhanced due to the return-current-induced decelerating electric field approaching 1GV/m. We have demonstrated the existence of collective effects, for high density beam, leading to enhanced stopping. This phenomenon will be important for the optimum design of ion driven inertial confinement fusion and fast ignition scenarios.

 \section{Method}
 We used the newly developed PIC code LAPINS \cite{lapins1,lapins2},where close interactions including proton-nuclei, proton-bound electron, proton-free electron were treated by a Monte Carlo binary collision model \cite{lapins4}. Furthermore a new Monte Carlo ionization dynamics model was added \cite{lapins3}, including collisional ionization, electron-ion recombination and ionization potential depression. Simulation of large scale plasmas often results in an intractable burden on computer power. Therefore, instead of solving the full Maxwell's equations, we used a new approach by combining the PIC method with a reduced model \cite{lapins1}. To take into account collective electromagnetic effects, the background electron inertia is neglected, and instead the background return current is evaluated by the Ampere's law $\textbf{J}_{e}$ = $(1/2\pi)\nabla\times\textbf{B}-(1/2\pi)\partial\textbf E /\partial t-\textbf J_b-\textbf J_i$, where $\textbf B$ is the magnetic field, $\textbf E$ is the electric field, and $\textbf J_i$ is the background ion current. Applying the continuity equation $\nabla\cdot\textbf J+\partial\rho/\partial t=0$ with the total current $\textbf J=\textbf J_b+\textbf J_i+\textbf J_e $, the Poisson Equation $\nabla\cdot\textbf E=2\pi\rho$ is rigorously satisfied. The electric fields are then obtained from Ohm's law, $\textbf E = \eta\textbf J_e-\textbf v_e\times\textbf B$, where $\textbf v_e$ is the background electron velocity, and $\eta$ is the resistivity. Taking advantage of the Monte Carlo collision model, resistivity $\eta$ is obtained by averaging over all binary collisions at each time step for each simulation cell. Finally, Faraday's law is used to obtain the magnetic fields $\partial\textbf B/\partial t = -\nabla \times \textbf E$. This field solver, which couples Ampere's law, Faraday's law and Ohm's law, can completely remove the numerical heating and reduces significantly the numerical expense. With these advantageous features a unique tool is at hand, which can self-consistently model transport and energy deposition of intense charged particles in dense ionized matter.

\bibliographystyle{unsrt}
\bibliography{ref}

\section{Aknowledgement}
We sincerely thank Olga Rosmej from GSI Helmholtzzentrum f$\ddot{u}$r Schwerionenforschung for the physical discussion, as well as the staff from Laser Fusion Research Center, Mianyang for the laser system running and target fabrication. The work is supported by Chinese Science Challenge Project No. TZ2016005, National Key Research and Development Project No. 2019YFA0404900, National Natural Science Foundation of China (Grant Numbers 11705141, 11775282, and U1532263), and China Postdoctoral Science Foundation (Grant Numbers 2017M623145 and 2018M643613). 

\section{Author contributions}
Yongtao Zhao conceived this work, organized the experiments and simulations with Weimin Zhou and Dong Wu, respectively. Jieru Ren, Zhigang Deng and Yongtao Zhao carried out the experiment together with the high power laser team (Zongqing Zhao, Weimin Zhou, and Yuqiu Gu), the plasma diagnostics team (Shaoyi Wang, Quanping Fan, Bubo Ma, Bo Cui, Xing Wang, Zhurong Cao and Leifeng Cao), ion beam characterization team (Wei Qi, Shuai Yin, Shukai He, Wei Liu, Rui Cheng, Xianming Zhou, Jianhua Feng, Yihang Zhang, and Zhe Zhang). Jieru Ren, Zhigang Deng, Shuai Yin, Wei Qi, Shaoyi Wang, and Quanping Fan analyzed the main part of the experimental data.  Benzheng Chen, Jieru Ren, Yongtao Zhao and Dong Wu performed the simulations and related analysis. Shaoping Zhu, Guoqing Xiao, Hongwei Zhao, Yutong Li, Yuqiu Gu and Leifeng Cao contribute in the physical discussion. Jieru Ren, Yongtao Zhao, Dong Wu and Dieter Hoffmann wrote the paper.

\section{Data availability}
The dataset generated and analyzed during the current study are available from the corresponding authors upon reasonable request. The simulation details are available from the corresponding author on reasonable request. 
\section{Additional information}
Supplementary information is available in the online version of the paper.
\section{Competing interest}
The authors declare no competing financial interests.

\end{document}